\documentclass[12pt]{article}
\usepackage{graphicx}
\usepackage{amssymb}
\usepackage{amscd}
\usepackage{amsmath}
\usepackage{cite}

\begin{document}
\begin{titlepage}

{\hbox to\hsize{\hfill January 2017 }}

\bigskip \vspace{3\baselineskip}

\begin{center}
{\bf \large 
Standard Model with hidden scale invariance and light dilaton}

\bigskip

\bigskip

{\bf Archil Kobakhidze and Shelley Liang \\ }

\smallskip

{ \small \it
ARC Centre of Excellence for Particle Physics at the Terascale, \\
School of Physics, The University of Sydney, NSW 2006, Australia, \\
E-mails: archil.kobakhidze, shelley.liang@sydney.edu.au
\\}

\bigskip
 
\bigskip

\bigskip

{\large \bf Abstract}

\end{center}
\noindent 
We consider the minimal Standard Model as an effective low-energy description of an unspecified fundamental theory with spontaneously broken conformal symmetry. The effective theory exhibits classical scale invariance which manifest itself through the dilaton field. The mass of the dilaton is generated via the quantum scale anomaly at two-loop level and is proportional to the techically stable hierarchy between the electroweak scale and a high energy scale given by a dilaton vacuum expectation value. We find that a generic prediction of this class of models is the existence of a very light dilaton with mass between $\sim 0.01$ $\mu$eV to $\sim 100$ MeV, depending on the hierarchy of scales. Searches for such a light scalar particle may reveal a fundamental role of conformal invariance in nature.      

 \end{titlepage}

\section{Introduction}

The discovery of the Higgs boson completes the Standard Model (SM) and confirms of the basic picture of mass generation through the spontaneous electroweak symmetry breaking. At the same time, the quadratic sensitivity of the Higgs mass under the quantum correction from ultraviolet physics and the related mass hierarchy problem remains a mystery. The measured Higgs boson mass, $m_{h}\simeq125$ GeV, can hardly be accommodated in the most popular minimal supersymmetric extension of the SM, which for a long time has been assumed as a prototype model for the solution of the hierarchy problem. 

As an alternative to supersymmetry, scale invariance has been advocated as a potential solution to the hierarchy problem for quite some time now \cite{Wetterich:1983bi, Bardeen:1995kv, Meissner:2006zh, Foot:2013hna, Aoki:2012xs, Kobakhidze:2014afa}. Conformal invariance and supersymmetry are believed to be symmetries of fundamental string theory. Conformal invariance is typically assumed to be broken at Planck/string scale, while supersymmetry survives all the way down the electroweak scale. As a logical possibility, one may also consider scenario where supersymmetry is broken at high energy scales, while conformal invariance is maintaining down to lower energy scales. In this paper we consider a low-energy effective description of such a scenario without trying to specify the ultraviolet completion. Spontaneously broken conformal invariance manifests in the effective theory through the dilaton field\footnote{Despite 5 broken generators, spontaneously broken conformal invariance in 4D results in a single (pseudo)Goldstone boson, the dilaton.}. More specifically we consider the minimal Standard Model with hidden scale invariance and demonstrate that technically natural hierarchy between the electroweak scale given by the Higgs vacuum expectation value (VEV) $v_{ew}\approx 246$ GeV and high energy scale $\Lambda$ defined by the dilaton VEV, $\epsilon =v_{ew}/\Lambda$, can be maintained in the effective theory. A rather generic prediction of our theory is the existence of light dilaton, which develops its mass via the dimensional transmutation mechanism due to the quantum scale anomaly. Assuming the vacuum energy is tuned to be (nearly) zero, as required by observations, the dilaton mass is generated at two-loop level in perturbation theory. It is also suppressed  as $\propto \epsilon$ and ranges between  $\sim 0.01$ $\mu$eV to $\sim 100$ MeV.

\section{The model}

There is little doubt that SM is an effective low energy description of some more fundamental theory, which incorporates dark matter and neutrino masses and perhaps also addresses other theoretical problems, such as strong CP problem and flavour problem as well as provides a framework for a consistent quantum description of gravity. As an effective theory SM contains an additional mass parameter, the ultraviolet cut-off $\Lambda$, which is not just a mathematical tool to regulate divergent amplitudes, but is a physical parameter that encapsulates physics (massive fields and high momenta modes of light fields) which we are agnostic of. The Higgs potential defined at this ultraviolet scale reads:
\begin{equation}
V(\Phi^{\dagger}\Phi)=V_0(\Lambda)+\lambda(\Lambda)\left[\Phi^{\dagger}\Phi - v_{ew}^2(\Lambda)\right]^2
+...,
\label{1}
\end{equation}       
where $\Phi$ is the electroweak doublet Higgs field, $V_0$ is the field-independent constant (bare cosmological constant parameter) and the ellipsis denote all possible dimension $> 4$ (irrelevant), gauge invariant operators, $\left(\Phi^{\dagger}\Phi\right)^n$, $n=3,4...$. Other bare parameters include dimensionless couplings $\lambda(\Lambda)$ and a mass dimension parameter $v_{ew}(\Lambda)$, the bare Higgs expectation value. In principle, this potential with infinite number of nonrenormalisable operators and $\Lambda$-dependent parameters must fully encode the physics beyond SM. In practice, however, the parameters are measured in low-energy experiments, which are not particularly sensitive to irrelevant operators. The truncated theory contains finite number of parameters and is reliable only in the low-energy domain. Now, if one computes quantum correction $\delta m_{\Phi}^2$  to the Higgs mass parameter $m_{\Phi}^2\equiv 2\lambda v^2_{ew}$ one finds that it is $\propto \Lambda^2$. Taking this computation as a guiding estimate, one comes to the conclusion that a light Higgs $(m_{\Phi}^2+\delta m_{\Phi}^2)/\Lambda^2 << 1$, necessarily implies fine-tuning between the tree-level parameter $m_{\Phi}^2(\Lambda)$ and the quantum correction to it, $\delta m_{\Phi}^2$.  However, this naive conclusion is not necessarily correct if the theory exhibits additional symmetries such as softly broken supersymmetry or classical scale invariance (for more discussion see Ref. \cite{Kobakhidze:2014afa}).    

Assume now that a fundamental theory maintains spontaneously broken scale invariance, such that all mass parameters (including gravitational constant) have the common origin. To make this symmetry manifest in our effective theory, we promote all mass parameters to a dynamical field $\chi$, the dilaton, as follows\footnote{See also earlier work \cite{Buchmuller:1990pz} for a similar inclusion of a dilaton within the dimensionally regularised SM. In a theory with gravity, the conformal scalar may play the role of the dilaton \cite{Kobakhidze:2015jya}.}:
\begin{equation}
\Lambda \to \Lambda \frac{\chi}{f_{\chi}},~~v_{ew}^2(\Lambda) \to \frac{v_{ew}^2(\chi)}{f_{\chi}^2}\chi^2\equiv \frac{\xi (\chi)}{2}\chi^2,~~V_0(\Lambda)\to \frac{V_0(\chi)}{f_{\chi}^4}\chi^4\equiv \frac{\rho(\chi)}{4}\chi^4~,  
\label{2}
\end{equation}  
where $f_{\chi}$ is the dilaton decay constant (in analogy with the pion decay constant in the effective chiral theory), which we assume to be equal to $\Lambda$ in what follows. Then, Eq. (\ref{1}) turns into the Higgs-dilaton potential, 
 \begin{equation}
V(\Phi^{\dagger}\Phi, \chi)=\lambda(\chi)\left[\Phi^{\dagger}\Phi  -\frac{\xi(\chi)}{2}\chi^2  \right]^2 +\frac{\rho(\chi)}{4}\chi^4~.  
\label{3}
\end{equation}  
This potential is manifestly scale invariant up to the quantum scale anomaly, which is engraved in $\chi$-dependence of dimensionless couplings\footnote{In this we differ substantially from the so-called quantum scale-invariant SM \cite{Shaposhnikov:2008xi, Ghilencea:2016dsl}. In their approach SM is extrapolated to an arbitrary high energy scale and regularized by invoking dilaton-dependent renormalization scale, $\mu=\mu(\chi)$}. Indeed, the Taylor expansion around an arbitrary fixed scale $\mu$ reads:
 \begin{equation}
\lambda^{(i)}(\chi)=\lambda^{(i)}(\mu)+\beta_{\lambda^{(i)}}(\mu)\ln\left(\chi/\mu\right)+\beta'_{\lambda^{(i)}}(\mu)\ln^2\left(\chi/\mu\right)+...,
\label{4}
\end{equation}  
where $\lambda^{(i)} \equiv (\lambda, \xi, \rho)$ and
 \begin{equation}
\beta_{\lambda^{(i)}}(\mu)=\left. \frac{\partial \lambda^{(i)}}{\partial \ln\chi}\right |_{\chi=\mu}~,
\label{5}
\end{equation}  
is the renormalisation group (RG) $\beta$-functions for respective couplings $\lambda^{(i)}$ defined at a scale $\mu$, while $\beta'_{\lambda^{(i)}}(\mu)=\left. \frac{\partial^2 \lambda^{(i)}}{\partial (\ln\chi )^2}\right |_{\chi=\mu}$, etc. Note that while the lowest order contribution in $\beta$-functions is one-loop, i.e. $\sim {\cal O}(\hbar)$, $n$-th derivative of $\beta$ is higher $nth$ order  in the perturbative loop expansion, $\sim {\cal O}(\hbar^n)$.  

In order to analyse minima of the potential (\ref{3}) it is convenient to set an arbitrary renormalisation scale $\mu$ to be equal to the dilaton VEV, $\langle \chi \rangle \equiv v_{\chi}$.  We also need to satisfy phenomenologically important constraint that 
the vacuum energy density is (nearly) zero $V(v_{ew}, v_(\chi))=0$ as it is required by astrophysical observations. The later constraint is nothing but a fine-tuning of the cosmological constant, which in scale invariant theories results in a certain relation between dimensioneless couplings \cite{Foot:2010et, Foot:2011et}. For our model we find:  
\begin{eqnarray}
V(v_{ew}, v_{\chi})=0 \Longrightarrow \rho(v_{\chi})=0~.
\label{6}
\end{eqnarray}  
This relation, together with the extremum condition $\left. \frac{dV}{d\chi}\right |_{\Phi=\langle\Phi\rangle, \chi=\langle\chi\rangle}=0$, actually implies:
 \begin{eqnarray}
\frac{\beta_{\rho}(v_{\chi})}{4}+\rho(v_{\chi})=0 \Longrightarrow \beta_{\rho}(v_{\chi})=0~.
\label{7}
\end{eqnarray}  
One of the above  equations (\ref{6}, \ref{7}) can be used to define the dilaton VEV (dimensional transmutation) and another represents tuning of the cosmological constant.   

The second extremum condition  $\left. \frac{dV}{d\Phi}\right |_{\Phi=\langle\Phi\rangle, \chi=\langle\chi\rangle}=0$ sets the hierarchy of VEVs:
 \begin{eqnarray}
\xi(v_{\chi})\equiv \epsilon^2 = \frac{v^2_{ew}}{v^2_{\chi}}~.
\label{8}
\end{eqnarray} 
This is a good place to remark on the the stability of the hierarchy of scales. Unlike the Higgs self-coupling $\lambda$, both  Higgs-dilatton and self-dilaton couplings, $\lambda \xi$ and $\lambda \xi^2 +\rho$, respectively, 
exhibit trivial infrared fixed points, i.e., $\xi=\rho=0$, and hence they do not change much under the RG running, if taken to be small at some renormalisation scale. This implies that the ratio of VEVs in Eq. (\ref{8}) can be hierarchically small in the sense of technical naturalness \cite{Wetterich:1983bi, Foot:2013hna, Kobakhidze:2014afa}, that is, no radiative corrections can change the hierarchy $\epsilon$ appreciably as it is defined through the small coupling $\xi$ according to Eq. (\ref{8}). We also note  that in the classical (or exact conformal) limit where $\beta_{\lambda^{(i)}}= 0$, VEVs and consequently the hierarchy 
 are undetermined and thus can be arbitrary, as expected.

\section{The light dilaton}
Next we compute the scalar mass spectrum.  The 2-by-2 mass squared matrix of the neutral Higgs scalar and the dilaton fields is given by  
  \begin{eqnarray}
\mathbf{M}^2(v_{\chi})= v_{ew}^2 \left(
\begin{tabular}{rr}
$2\lambda(v_{\chi})$  &  $-\frac{\lambda(v_{\chi})}{\epsilon}\left(\beta_{\xi}(v_{\chi})  + 2\epsilon^2 \right)$ \\ 
 $-\frac{\lambda(v_{\chi})}{\epsilon}\left(\beta_{\xi}(v_{\chi})  + 2\epsilon^2 \right)$  & 
$\frac{\lambda(v_{\chi})}{2\epsilon^2}\left(\beta_{\xi}(v_{\chi})  + 2\epsilon^2 \right)^2+\frac{\beta'_{\rho}(v_{\chi})}{4\epsilon^2}$
 \\ 
\end{tabular} 
\right)
\label{11}
\end{eqnarray}  
We immediately notice that in the limit $\beta'_{\rho}(v_{\chi})\to 0$ the above matrix becomes degenerate and hence the dilaton running mass tends to zero at the scale $v_{\chi}$\footnote{Note however, very small mass is still expected due to the RG running in the infrared. Obviously, the dilaton would be strictly massless in the full conformal limit, as it is a true Goldstone boson in this limit.}. Thus the dilaton mass in our model emerges at ${\cal O}(\hbar ^2)$ in the perturbative loop expansion. This is in accord with the earlier observation in \cite{Foot:2011et} that cancellation of the scalar vacuum energy implies that the dilaton mass is generated at 2-loop level. More specifically, we find for the scalar running masses,   
\begin{eqnarray}
m_h^2\simeq2\lambda (v_{\chi}) v^2_{ew}~,~~ m_{\chi}^2\simeq\frac{\beta'_{\rho}(v_{\chi})}{4\epsilon^2}v^2_{ew}~,
\label{12}
\end{eqnarray} 
where, to a good accuracy, we can express $\beta'_{\rho}(v_{\chi})$ through the Higgs self-coupling beta-function as:
\begin{eqnarray}
\beta'_{\rho}(v_{\chi})= \epsilon^4\beta'_{\lambda}(v_{\chi})+2\epsilon^2\beta'_{\lambda_{h\chi}}(v_{\chi})+\beta'_{\lambda_{\chi}}(v_{\chi})-\frac{2}{\lambda(v_{\chi})}\left(\epsilon^2\beta_{\lambda}(v_{\chi})+\beta_{\lambda_{h\chi}}(v_{\chi})\right)^2~,
\label{13}
\end{eqnarray}
where the relevant beta-functions are given in appendix. We observe that the Higgs boson mass in Eq. (\ref{12})  is essentially the same as in SM, while dilaton mass is suppressed by the hierarchy parameter, $m_{\chi}\propto \frac{\epsilon}{16\pi^2} v_{ew}$. We also find that Higgs-dilaton mixing is small and is also controlled by the hierarchy parameter:
 \begin{eqnarray}
\tan2\alpha \approx -\epsilon~,
\label{14}
\end{eqnarray} 
,e.g. $\alpha \lesssim 0.01$ for $v_{\chi}\gtrsim 10$ TeV.  Thus, our model predicts a very light dilaton  with very small mixing with the Higgs boson. 

\begin{figure}[t] 
\includegraphics[width=0.5\linewidth]{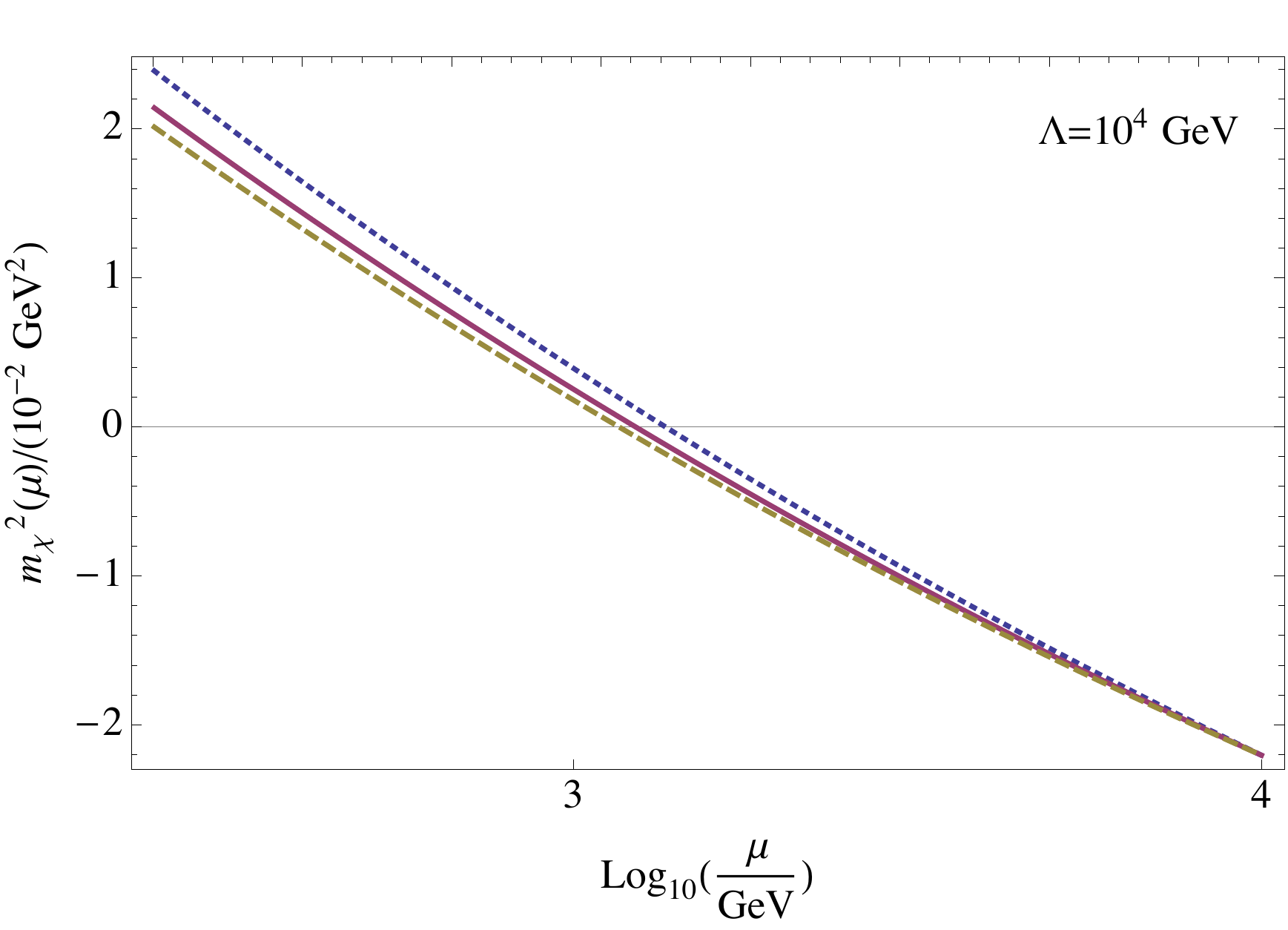} 
\includegraphics[width=0.5\linewidth]{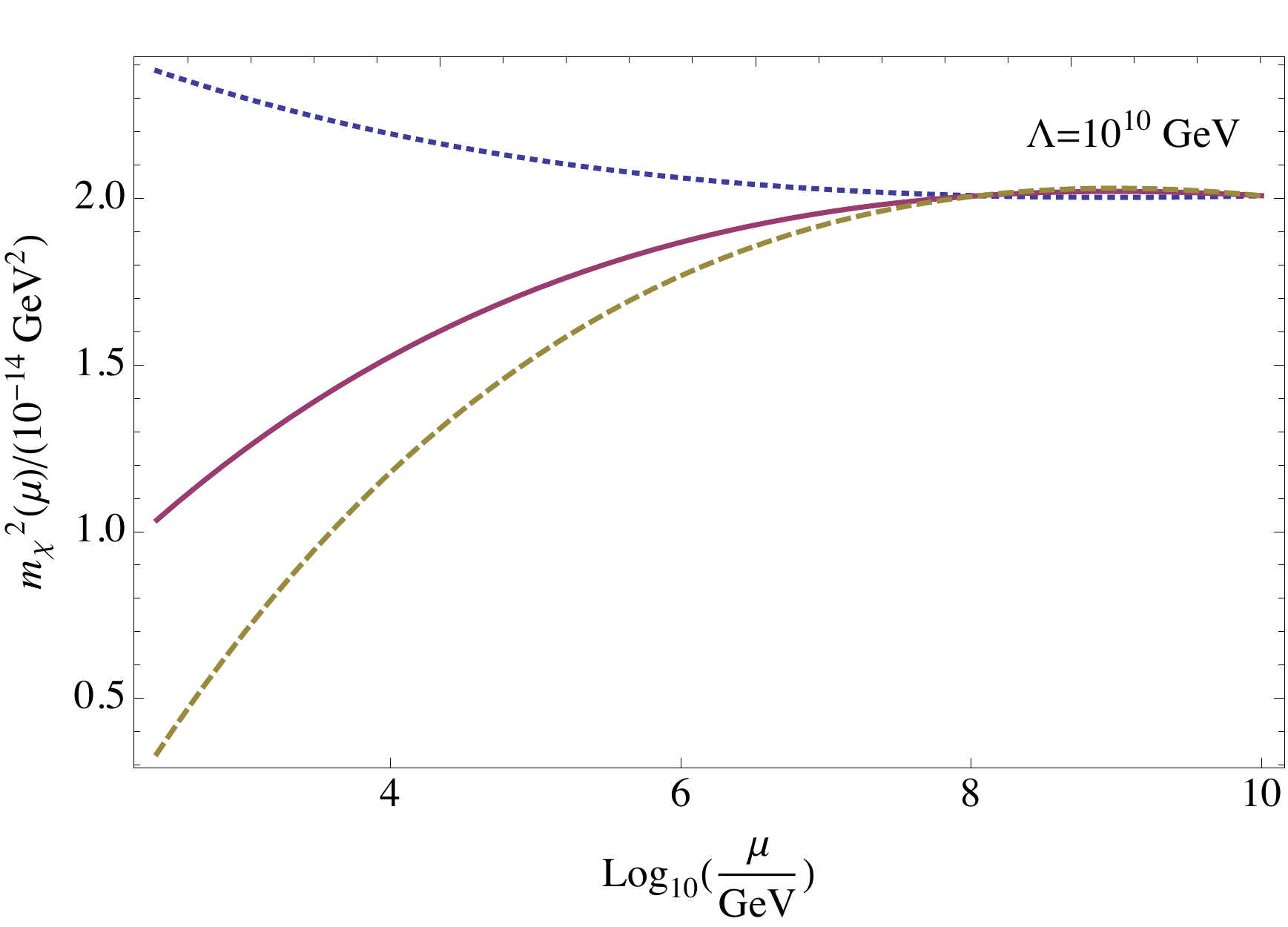} 
\includegraphics[width=0.5\linewidth]{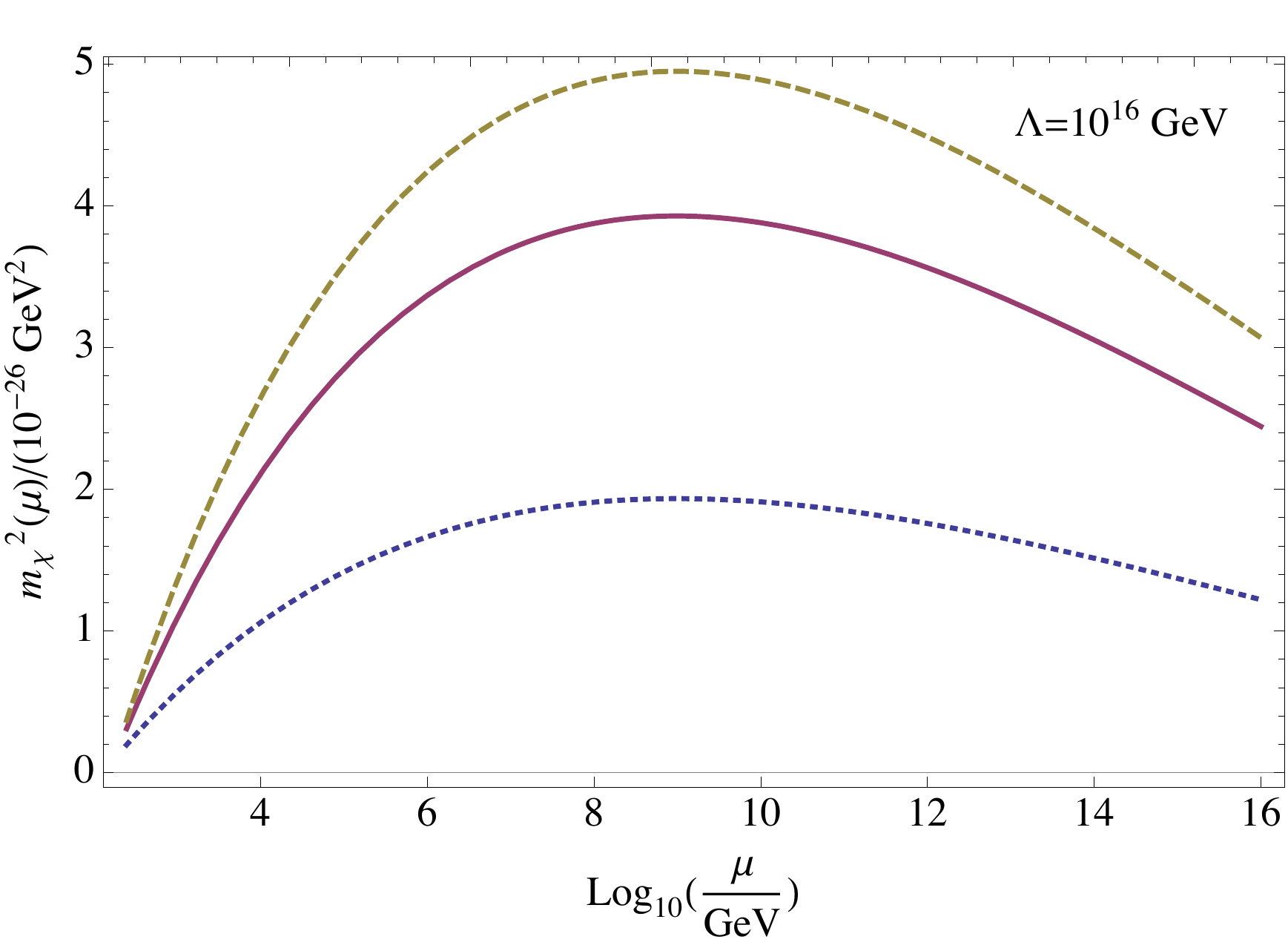} 
\includegraphics[width=0.5\linewidth]{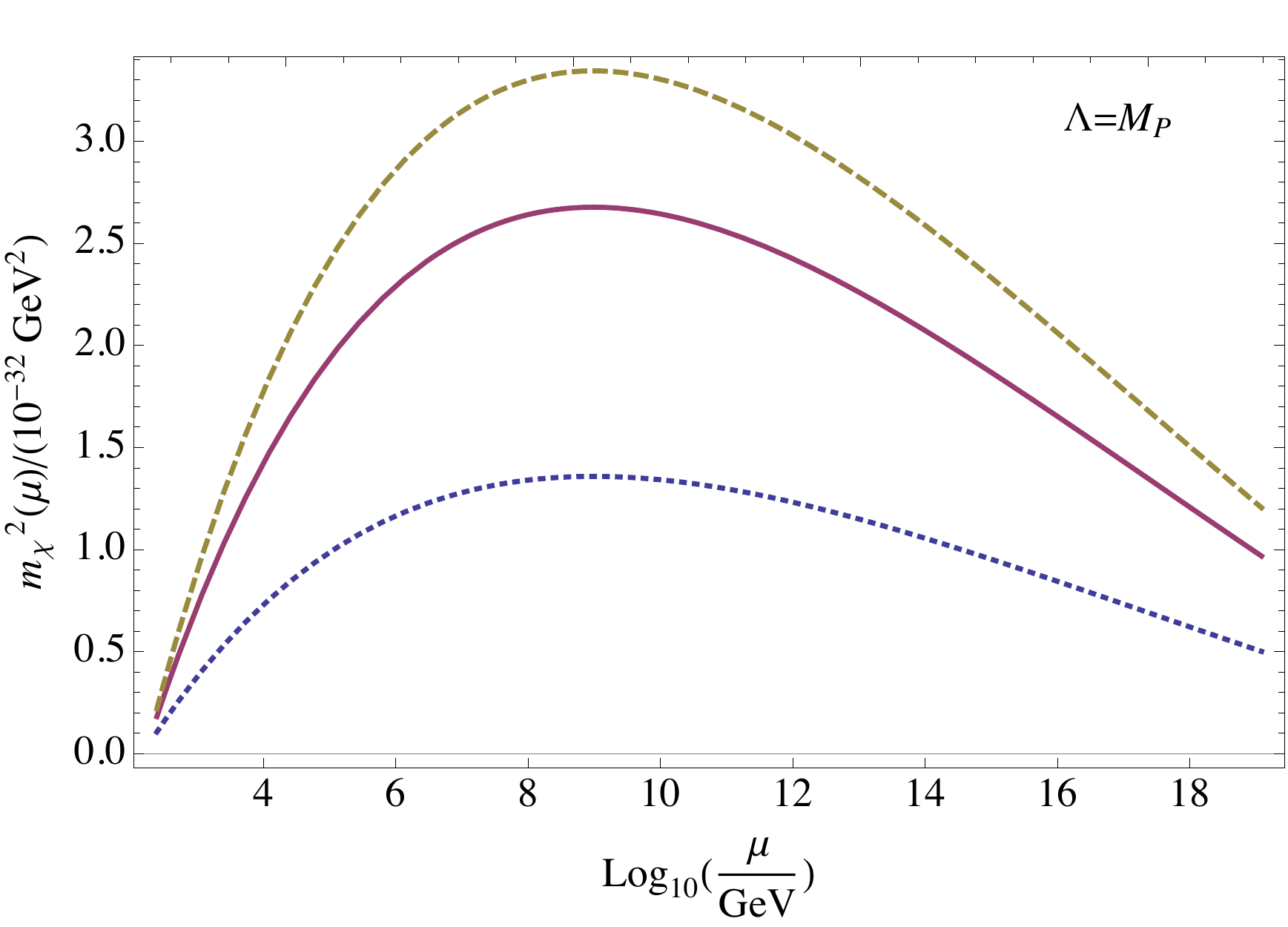} 
\caption{\small RG evolution of the dilaton mass square for various cut-off scales: $10^4,~10^{10},~10^{16}$ and $M_P= 1.2\cdot 10^{19}$ GeV. The dotted, solid and dashed lines corresponds to top quark mass $m_t= 171, 173$ and 174 GeV, respectively.} 
\label{mass}
\end{figure}

The sign of the running masses in Eq. (\ref{12}) at the cut-off scale (recall $v_{\chi}=\Lambda$) is essentially defined by the largest scalar coupling $\lambda (\Lambda)$. The RG evolution of this coupling (and thus its $\Lambda$-dependence) in our model  is very similar to the one in SM:   $\lambda (\mu)$ becomes negative at a scale $\mu_I\sim 10^{8}$ GeV, signalling instability of the effective potential\footnote{It is known that the electroweak vacuum in SM is a metastable state (see the most recent analysis in Refs \cite{Bednyakov:2015sca, Espinosa:2015qea}) and is consistent with observations, unless the rate of inflation is large \cite{Kobakhidze:2013tn, Espinosa:2015qea}. The potential instability due to the fast inflation, however, must be re-analysed in our model, since dilaton is expected to play a significant role.}. We find that the dilaton mass square $m^2_{\chi}(\Lambda)$ is positive for negative $\lambda(\Lambda)$, i.e. for $\mu_I\lesssim\Lambda \lesssim 10^{17}$ GeV, and negative when  $\lambda(\Lambda)>0$. However, evaluating this running mass down to the infrared region, we find that it is always positive and well approximated by the formulae: $m_{\chi}\propto \frac{\epsilon}{16\pi^2} v_{ew}$. Hence depending on cut-off $\Lambda \in \left[10^4~ {\rm GeV},~~10^{19}~ {\rm GeV} \right]$, we predict light dilaton with mass in the range from $\sim 0.01$ $\mu$eV to $\sim 100$ MeV. The results of these calculations are presented on Figure \ref{mass}. 

The couplings of the dilaton with the SM particles are defined through the mixing with the Higgs boson and scale anomaly. Since the dilaton is a pseudo-Goldstone boson of spontaneously broken anomalous scale invariance, its couplings are suppressed by powers of $1/v_{\chi}$. For large $v_{\chi}$, the dilaton is a very light state which feebly interacts with SM particles and can, in principle, play the role of dark matter. We will study phenomenology of the light dilaton in details elsewhere.     

\section{Conclusion}
In this paper we have presented a very simple extension of the effective SM with hidden scale invariance. The scale invariance manifests at low energies through the dilaton field. All mass scales in the model (including Wilsonian cut-off of the effective theory) are generated through the dilaton VEV $v_{\chi}$ through the quantum effect of dimensional transmutation. We have argued that $v_{ew}/v_{\chi}\ll 1$ is technically natural. In addition, assuming cancellation of the scalar vacuum energy, the dilaton mass is generated at 2-loop level, $\sim {\cal O}(\hbar^2)$, in the perturbative loop expansion and is proportional to the hierarchy of scales $\epsilon=v_{ew}/v_{\chi}$. Therefore, light dilaton with mass between $\sim 0.01$ $\mu$eV to $\sim 100$ MeV, depending on $\epsilon$, is a generic prediction of our model.  Searches for such a light scalar particle may reveal a fundamental role of conformal invariance in nature.

Besides the phenomenological studies of the light dilaton, which we delegate to future work, an interesting extension of our present work would be construction of a model which also addresses other outstanding problems of SM, such as neutrino masses and the strong CP problem. Inclusion of gravity in the current framework and study of early universe models of electroweak phase transition \cite{new} or inflationary scenarios along the lines of Ref. \cite{Barrie:2016rnv} would also be interesting.   

\paragraph{Acknowledgement.} The work was supported in part by the Australian Research Council.

\appendix
\section{Beta functions}

For reader's convenience here we include the relevant one-loop beta-functions, $\beta_{C}=\frac{dC}{d\ln(\mu)}$, 
used in our calculations:
\begin{eqnarray*}
\beta_{g_{Y}} & = & \frac{g_{Y}^{3}}{16\pi^{2}}\frac{41}{6},\quad\beta_{g_{2}}=\frac{g_{2}^{3}}{16\pi^{2}}\left(-\frac{19}{6}\right),\quad\beta_{g_{3}}=\frac{g_{3}^{3}}{16\pi^{2}}\left(-7\right)\\
\beta_{y_{t}} & = & \frac{y_{t}}{16\pi^{2}}\left(-\frac{9}{4}g_{2}^{2}-8g_{3}^{2}-\frac{17}{12}g_{Y}^{2}+\frac{9}{2}y_{t}^{2}\right)\\
\beta_{\lambda} & = & \frac{1}{16\pi^2}\left[\lambda\left(-9g_{2}^{2}-3g_{Y}^{2}+24\lambda+12y_{t}^{2}\right)+\frac{3}{4}g_{2}^{2}g_{Y}^{2}+\frac{9}{8}g_{2}^{4}+\frac{3}{8}g_{Y}^{4}-6y_{t}^{4}+2\lambda_{h\chi}^{2}\right]\\
\beta_{\lambda_{\chi}} & = & \frac{1}{16\pi^{2}}\left(18\lambda_{\chi}+8\lambda_{h\chi}^{2}\right)\\
\beta_{\lambda_{h\chi}} & = & \frac{\lambda_{h\chi}}{16\pi^2}\left(-\frac{9}{2}g_{2}^{2}-\frac{3}{2}g_{Y}^{2}+12\lambda+6\lambda_{\chi}+8\lambda_{h\chi}+6y_{t}^{2}\right)\\
\beta_{m_{\chi}^{2}} & = & \frac{v^{2}}{16\pi^{2}}8\lambda\lambda_{h\chi}
\end{eqnarray*}

The RG equations were solved using the values of couplings at $\mu=m_t$ and relations between couplings at $\mu=v_{\chi}$ steaming from minimization conditions as described in the main text:
\begin{eqnarray*}
\lambda(m_{t}) & = & \frac{1}{2}\frac{m_{h}^{2}}{v_{ew}^{2}}\approx 0.129\\
g_{Y}(m_{t}) & = & 0.35761+0.00011\left(\frac{m_{t}}{\text{GeV}}-173.10\right)\\
g_{2}(m_{t}) & = & 0.64822+0.00004\left(\frac{m_{t}}{\text{GeV}}-173.10\right)\\
g_{3}(m_{t}) & = & 1.1666-0.00046\left(\frac{m_{t}}{\text{GeV}}-173.10\right)+\frac{0.00314\left(\alpha_{3}(m_{Z})-0.1184\right)}{0.0007}\\
y_{t}(m_{t}) & = & 0.93558+0.0055\left(\frac{m_{t}}{\text{GeV}}-173.10\right)-\frac{0.00042(\alpha_{3}(m_{Z})-0.1184)}{0.0007}\\
\alpha_{3}(m_{Z}) & = & 0.1185\pm0.0006,
\end{eqnarray*}
where $m_{t}$ is the pole mass of top quark and the Higgs mass is taken to be $m_{h}=125.09$ GeV.

In addition, the following relations, steaming from the minimization of the scalar potential, must hold at the cut-off scale $\Lambda=v_{\chi}$:
\begin{equation*}
\lambda_{\chi}(v_{\chi})=-\epsilon^2\lambda_{h\chi}(v_{\chi})=\epsilon^4\lambda(v_{\chi})~.
\end{equation*}

\end{document}